 \documentclass[preprint,12pt]{elsarticle}

\usepackage{subfigure,dcolumn}
\usepackage{epstopdf}
\usepackage{mhchem}
\usepackage{upgreek}
\usepackage{graphicx}
\usepackage{subfigure}
\usepackage{multirow}
\usepackage{footnote}
\usepackage{setspace}
\usepackage{ulem} 
\usepackage[dvipsnames]{xcolor}
\usepackage{physics}
\usepackage{xparse}
\usepackage{amsmath}
\usepackage{cuted}
\usepackage{color}
\usepackage{lineno}

\usepackage{float}
\usepackage{stfloats}
\usepackage{graphicx}
\usepackage{booktabs} 
\usepackage{setspace}
\usepackage{tabularx}
\usepackage[colorlinks,
            linkcolor=blue,
            anchorcolor=blue,
            citecolor=blue]{hyperref}

\usepackage{caption} 
\usepackage{flushend}
\usepackage{appendix}

\journal{Nuclear Instruments and Methods in Physics Research A}

\begin{document}
\begin{frontmatter}

\title{Track Recognition for the $\Delta E-E$ Telescopes with Silicon Strip Detectors}

\author[THU]{Fenhai Guan\corref{cor1}} 
\author[THU]{Yijie Wang}  
\author[THU]{Xinyue Diao} 
\author[THU]{Yuhao Qin}   
\author[THU]{Zhi Qin}     
\author[THU]{Dong Guo}    
\author[THU]{Qianghua Wu} 
\author[THU]{Dawei Si}    
\author[THU]{Sheng Xiao}  
\author[THU]{Boyuan Zhang}  
\author[THU]{Yaopeng Zhang} 
\author[THU]{Xuan Zhao}     
\author[THU]{Zhigang Xiao}  
\address[THU]{Department of Physics, Tsinghua University, Beijing 100084, China}
\cortext[cor1]{gfh16@mails.tsinghua.edu.cn (Corr. author)}

\begin{abstract}

For the high granularity and high energy resolution, Silicon Strip Detector (SSD) is widely applied in assembling telescopes to measure the charged particles in heavy ion reactions. In this paper, we present a novel method to achieve track recognition in the SSD telescopes of  the Compact Spectrometer for Heavy Ion Experiment (CSHINE).  Each telescope consists of a single-sided silicon strip detector (SSSSD) and a double-sided silicon strip detector (DSSSD) backed by $3 \times 3$ CsI(Tl) crystals. Detector calibration and track reconstruction are implemented. Special decoding algorithm is developed for the multi-track recognition procedure to deal with the multi-hit effect convoluted by charge sharing and the missing signals with certain probability.  It is demonstrated that the track recognition efficiency of the method is approximately 90\% and 80\% for the DSSSD-CsI and SSSSD-DSSSD events, respectively.

\end{abstract}


\end{frontmatter}


\section{Introduction}\label{sec. I}

In recent years, the $\Delta E-E$ telescopes with silicon strip detectors (SSDs) or pixel detectors are widely used in nuclear reaction experiments because of the excellent energy and position resolution as well as the good particle identification capability. In 
terrestrial laboratories, many such large acceptance detector systems including INDRA ~\cite{indra95,indra18}, LASSA ~\cite{lassa01}, HiRA ~\cite{hira07}, CHIMERA ~\cite{chimera02}, MUST2 ~\cite{must05}, FARCOS ~\cite{farcos13,farcos16}, FAZIA ~\cite{fazia16}, and ChAKRA~\cite{chakra19}, etc, have been developed for the studies of  nuclear reactions, nuclear structure and particle-particle correlation. In order to investigate the nuclear equation of state (nEoS), the HBT correlation, as well as the fast fission following heavy ion reactions (HIRs)~\cite{wuqh19,wuqh20}, a compact spectrometer for heavy ion experiments  (CSHINE) in the Fermi energy regime has been recently built~\cite{wyjnst21,gfh21}. CSHINE at the current stage consists of four telescopes and three large-area parallel plate avalanche counters (PPACs). The PPACs are used to measure the fission fragments while the telescopes are used to detect the light charged particles (LCPs). Each CSHINE  telescope consists of a single-sided silicon strip detector (SSSSD) and a double-sided silicon strip detector (DSSSD) backed by $3 \times 3$ CsI(Tl) crystals.  With the SSD-SSD-CsI configuration, particle identification in a wide energy range can be achieved by the $\Delta E-E$ correlation of DSSSD-CsI and SSSSD-DSSSD.

Before conducting physical analysis, we have to reconstruct the physical event first, which mainly includes energy calibration of the detectors, particle identification and track reconstruction. In principle, the position of an incident  particle is determined by the DSSSD in the unit of ``pixel", the size of which is given by the strip width. However, the task of track reconstruction is somewhat complicated because of  charge sharing and multi-hit effects. 

In this work, we present a novel method for event reconstruction of the SSD-SSD-CsI telescopes. The paper is arranged as following. Section \ref{sec. II} presents the telescope structure and the detector calibration. Section \ref{sec. III} represents the track reconstruction. Section \ref{sec. V} represents the results of the phase space distribution of the LCPs and Section \ref{sec. VI} is the summary.

\section{Detector calibration} \label{sec. II}

Both silicon strip detectors and the CsI hodoscope require calibration prior to physical analysis. While the SSD is calibrated using the conventional source plus pulser method, the CsI calibration relys on the particle species obtained from the $\Delta E-E$ plot. Before the calibration of CsI units, particle identification (PID) on $\Delta E-E$ plot has to be established.  Thus,  the calibration procedure is convoluted with the track recognition in the SSD telescope. This section presents the procedure of the detector calibration .

\subsection{Telescope configuration}\label{sec. II0}
\begin{figure}[t] 
  \centering
  \includegraphics[width=0.55\textwidth]{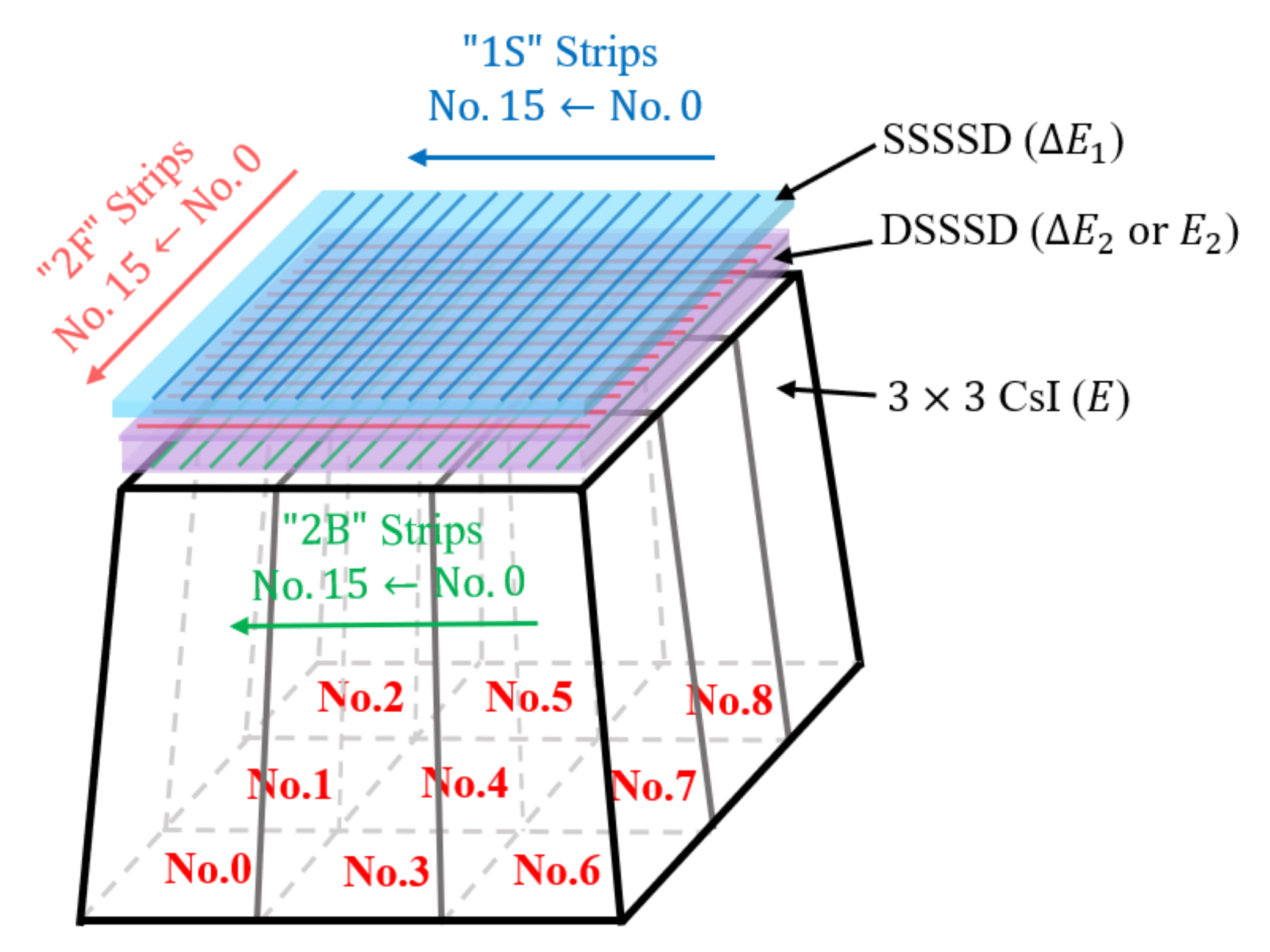} 
  \caption{ (Color online) The schematic diagram of a single CSHINE telescope. Each telescope consists of a SSSSD for $\Delta E_1$ detection, a DSSSD for $\Delta E_2$ or $E_2$ detection and $3 \times 3$ CsI crystals for $E$ detection.}
  \label{fig:config}
\end{figure}

Fig. \ref{fig:config} shows the schematic diagram of a single CSHINE telescope. Each telescope is composed of a SSSSD, a DSSSD and a hodoscope of  $3 \times 3$ CsI(Tl) crystals. The SSDs are BB7 series with active areas of $63.8 \times 63.8$ $\rm{mm}^2$. On each SSD surface, there are 32 strips with the width of 2 mm, and the gap between adjacent strips is 100 $\mu$m. The strips on the SSSSD surface (``1S") and on the back-side of the DSSSD (``2B") are parallel to each other and are numbered in the same order, while the ``2B" strips are perpendicular to the front strips of the DSSSD (``2F"). In the recent experiment, every two strips are merged into one channel to reduce electronics, so there are 16 channels on each detector surface and the size of the pixel is determined to be $4 \times 4$ $\rm{{mm}^2}$. Nine trapezoidal crystals, each of which is 50 mm long, $\rm{23 \times 23}$ $\rm{{mm}^2}$ on the front and $\rm{27 \times 27}$ $\rm{{mm}^2}$ on the rear, are closely packed into one trapezoid hodoscope. The light from the crystal is readout by an $\rm{18 \times 18}$ $\rm{{mm}^2}$ S3204 photodiode (Hamamatsu), and transferred to a Mesytec MPR-16 preamplifier and a MSCF-16 amplifier successively for amplification and shaping. With such SSD-SSD-CsI configuration, we can detect LCPs in a wide energy range. The analysis in this work is based on the experimental data from ${\rm ^{86}Kr+^{208}Pb}$ reactions at 25 MeV/u. Table \ref{cshine-para} presents the parameters of the four SSD telescopes. More details can be found in Ref. \cite{gfh21}.

\begin{table}[tp]
  \caption{The parameters of the four CSHINE SSD telescops. $L$ is the distance from the detector to the target, $\theta_{\rm lab}$ and $\phi_{\rm lab}$ are polar angle and  azimuthal angle of the detector center in the laboratory frame, respectively.}
  \label{cshine-para}
  \setlength\tabcolsep{8pt}
  \centering
  \small
  \begin{tabular}{lccccc}
  \toprule
  Telescope No.  & 1   & 2  & 3  & 4 \\[1.5pt] 
  \midrule
  $L$ (mm)              & 315.5 & 275.5 & 275.5 & 215.5 \\[1.5pt]
  $\theta_{\rm lab}$ $(^{\circ})$ & 18    & 25    & 31    & 51    \\[1.5pt]
  $\phi_{\rm lab}$ $(^{\circ})$   & 302   & 218   & 126   & 81    \\[1.5pt]
  SSSSD ($\mu \rm{m}$)  & 304   & 305   & 110   & 70    \\[1.5pt]
  DSSSD ($\mu \rm{m}$)  & 1010  & 1008  & 526   & 306   \\[1.5pt]
  CsI(Tl) ($\rm{mm}$)   & 50    & 50    & 50    & 50    \\[1.5pt]
  \bottomrule
  \end{tabular}
\end{table}

\subsection{Energy calibration of the silicon strip detectors}\label{sec. II1}
Energy deposition ($\epsilon$) of charged particles in silicon detectors is independent on particle species, as long as the particles with $Z\le 6$ is concerned. 
In pulse calibration, we obtain the parameters $k^{\prime}$ and $b^{\prime}$ of the following functions between the input pulse amplitude ($V$) and ADC channels ($ch$),  
\begin{subequations}
\begin{align}
V &= k^{\prime} \times {ch} + b^{\prime} \\
\epsilon &= \frac{k^{\prime}}{\beta} \times {ch} + \frac{b^{\prime}}{\beta}
\label{eq:cali}
\end{align}
\end{subequations}
Assuming that $V$ is equivalent to $\epsilon$ by $V = \beta \epsilon$, then $\beta$ is determined by an $\alpha$ energy point, e.g.,  $\beta = (k^{\prime} \times {ch}(\alpha) + b^{\prime})/\epsilon(\alpha)$. The energy calibration  is finalized as (\ref{eq:cali}).

\begin{figure*}[t] 
  \centering
  \includegraphics[width=0.95\textwidth]{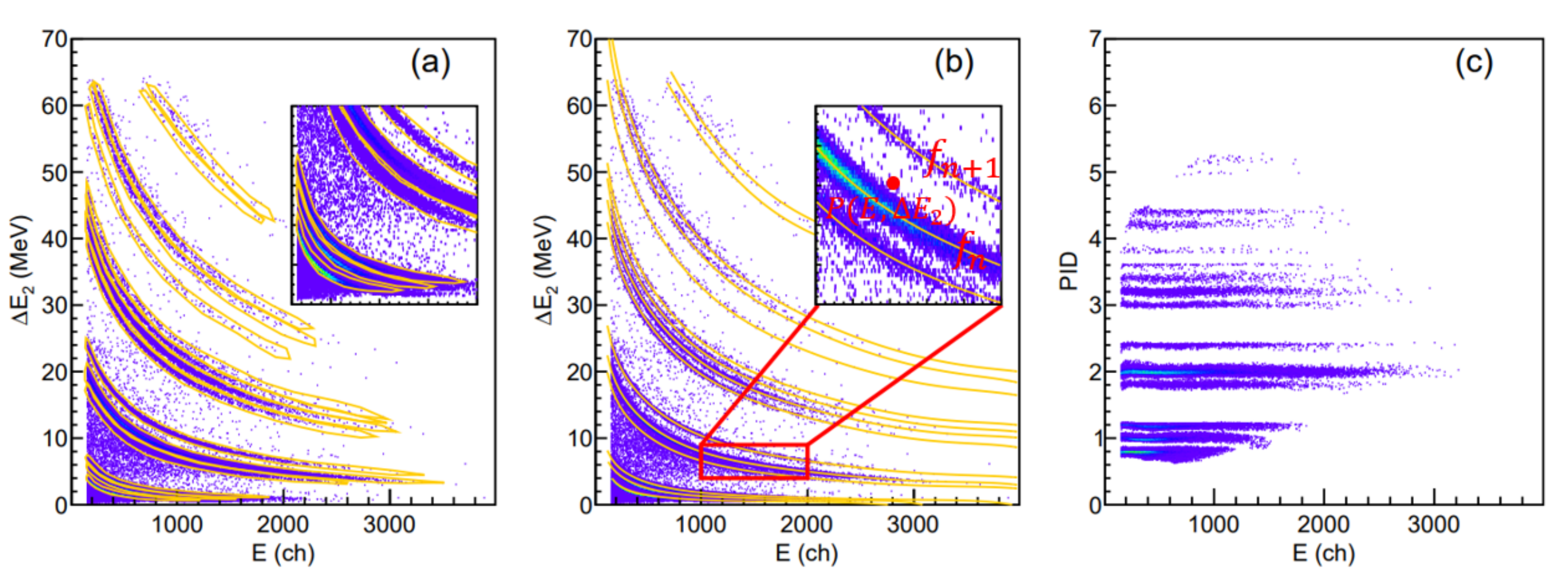} 
  \caption{(Color online) Particle identification of telescope No. 4. Panal (a) shows the inclusion contours for all the isotopes of a $\Delta E_{2}-E$ scattering plot, where $\Delta E_{2}$ and $E$ are detected energies in the DSSSD and in the CsI(Tl) crystal, respectively. Panel (b) depicts the fitting curves of the isotopes, and the inset shows the principle of the second PID method. Panel (c) shows the results of particle identification with isotope cuts shown in panel (a).} 
  \label{fig:pid}
\end{figure*}

\subsection{Particle identification with $\Delta E-E$ method}\label{sec. II2}
\subsubsection{Particle identification}\label{sec: II21}
Particle identification (PID) is of crucial importance in the detection of charged particles. Here we recall two conventional methods  used to achieve PID, which both rely on the $\Delta E-E$ correlation. For one method, each isotope is recognized by the inclusive  contour drawn $a~ priori$, as shown in Fig. \ref{fig:pid} (a).  For the other method, PID is achieved  for each particle on the $\Delta E-E$ scattering plot by computing the  distance from each individual point to  the neighboring ridge curves following the isotopes, as shown in Fig. \ref{fig:pid} (b).
It is explained as following. 

First, one fits the selected date points in the center of each isotope band with (\ref{polfit}),
\begin{equation} \tag{2a}
\begin{aligned}
f_{n}(x) =  a_{n}^{0}\cdot x^{-1} &+ a_{n}^{1}\cdot x^{0} + a_{n}^{2}\cdot x^{1} + a_{n}^{3}\cdot x^{2} + a_{n}^{4}\cdot x^{3}\\ &+ a_{n}^{5}\cdot x^{4} + a_{n}^{6}\cdot x^{5} + a_{n}^{7}\cdot x^{6} \label{polfit}  
\end{aligned}
\end{equation}
where $a_{n}^{0} \sim a_{n}^{7}$ are eight parameters, and $n$ represents different particles. Then, one defines the PID variable of the $n^{\rm th}$ particle as ~\cite{tassan02}
\begin{equation} \tag{2b}
\begin{aligned}
{\rm PID}_{n} = Z_{n} + 0.2(A_{n} - 2Z_{n}) \label{pid}
\end{aligned}
\end{equation}
where $Z_{n}$ and $A_{n}$ are integers referring to the charge and the mass numbers of the $n^{\rm th}$ fitting function or inclusion contour, respectively. Finally, we can define the PID variable of each particle as (\ref{pidexp}).
\begin{equation} \tag{2c}
\begin{aligned}
{\rm PID}(E,\Delta E_{2}) &= {\rm PID}_{n}  \\ &+ \frac{\Delta E_{2} - f_{n}(E)}{f_{n+1}(E) - f_{n}(E)} \cdot \left( {\rm PID}_{n+1} - {\rm PID}_{n} \right) \label{pidexp}
\end{aligned}
\end{equation}

As shown in Fig. \ref{fig:pid}(c), the isotope bands are straightened with (\ref{pid}) and (\ref{pidexp}). In addition, it is worth mentioning that Bethe formula has also been used to fit the ridge curves simultaneously for particle identification ~\cite{tassan02, lnsreport05}.

\begin{figure*}[t] 
  \centering
  \includegraphics[width=0.95\textwidth]{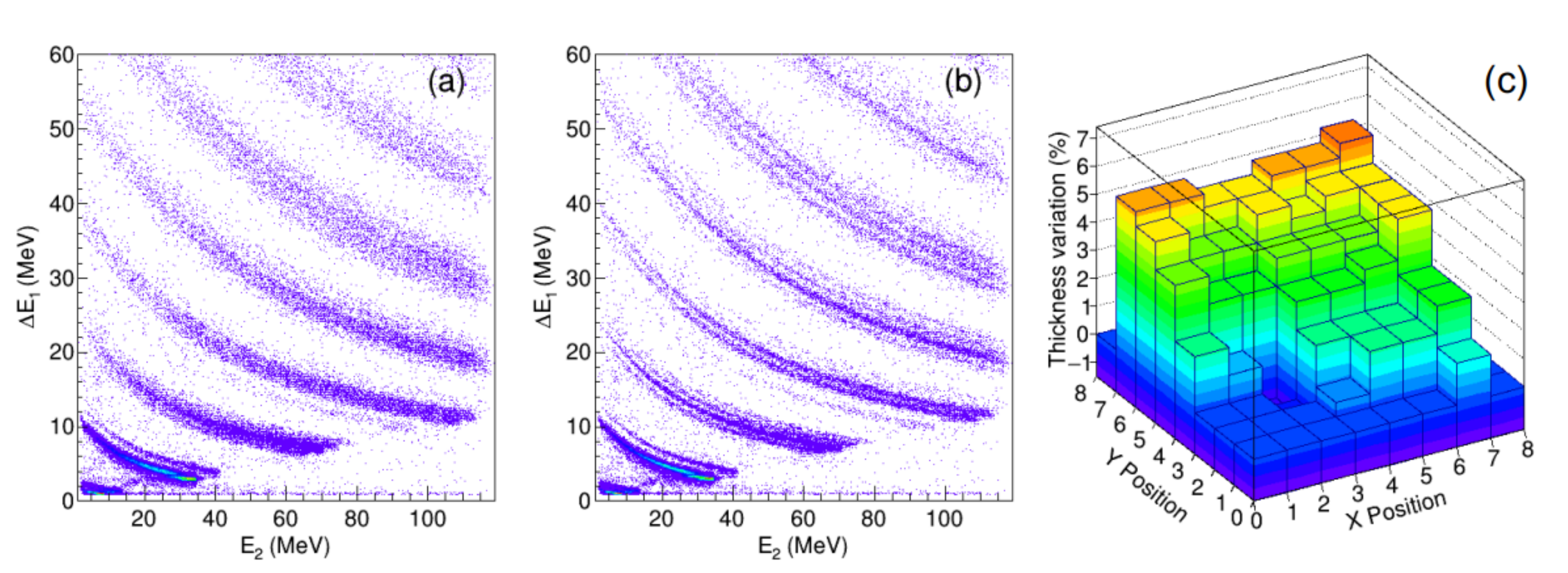} 
  \caption{(Color online) Panel (a) is a $\Delta E_{1}-E_{2}$ scattering plot before thickness correction for the 110 ${\rm \mu m}$ $\Delta E_{1}$ detector (the thickness of $E_{2}$  detector is 526 $\rm{\mu m}$). Panel (b) shows the 2D map after thickness correction for the $\Delta E_{1}$ detector. Panel (c) depicts the "relative" thickness correction for the $\Delta E_{1}$ detector.}
  \label{fig:thickness}
\end{figure*}

\subsubsection{Thickness correction for the thin silicon strip detector}\label{sec: II22}
The energy loss of a particle traversing a thin silicon detector is approximately proportional to the detector thickness. It is important to correct the thickness non-uniformity of very thin silicon strip detectors, in order to improve the identification resolution with $\Delta E-E$ technique. Here, we make ``relative" thickness correction based on the experimental data with particles penetrating the thin SSSSD and stopped in the following DSSSD.

For simplicity, we first select events with only one particle hitting both detectors. Considering the variation of the uniformity is smooth, the sensitive area is divided into $8 \times 8$ bins. In each bin, we draw the $\Delta E_{1}-E_{2}$ histogram individually and pick a certain bin, which exhibits clear PID bands, as a reference. Then, for the rest bins, the thickness correction factor $\eta$ for the SSSSD is varied from -0.05 to 0.10 with a step of 0.005, and energy loss $\Delta E_{1}$  in the SSSSD is corrected accordingly to be $\Delta E_{1} (1+\eta)$. For each $\eta$ value, a 2D histogram is filled and compared to the histogram of the reference bin. Finally, the thickness correction for each bin is determined with the $\eta$ value if the corresponding histogram holds the best consistence with the referenced one. Fig. \ref{fig:thickness} (b) shows the 2D plot after the thickness correction for the 110 $\rm{\mu m}$ detector, where the identification resolution is significantly improved when compared with the plot before correction as shown in Fig. \ref{fig:thickness} (a). Fig. \ref{fig:thickness} (c) is the ``relative" thickness variation for the 110 $\rm{\mu m}$ SSSSD,  which is in the range from $-1\%$ to $6\%$. This method provides the relative correlation of the thickness of a thin $\Delta E$ detector with respect to a referenced bin. In order to get the absolute thickness correction, one should rely on in-beam tests, as introduced in Refs.~\cite{lassa01,yyl18}.

\subsection{Energy calibration of the CsI(Tl) crystals}\label{sec. II3}
Energy calibration of the CsI(Tl) crystal is more complicated than that of the silicon since the light response of the CsI(Tl) crystal is non-linear and depends on the charge and mass of the  particles as well as on the length of the crystal. The calibration of CsI(Tl) crystal can be carried out by using the $\Delta E-E$ method if the forward DSSSD with a well-defined thickness is calibrated. For particles penetrating the DSSSD (with energy loss $\Delta E_{2}$) and stopped in the CsI(Tl) crystal, one can extract a set of data points on the ridge of each isotope band in particle identification, as shown in Fig. \ref{fig:pid}. The kinetic energy of the incident particle $E_{\rm k}$ corresponding to each selected point can be deduced by a numerical inversion of Ziegler's energy loss tables~\cite{ziegler} with $\Delta E_{2}$ (MeV) known in DSSSD. The residue energy in the CsI(Tl) crystal can be calculated as $E ({\rm{MeV}}) = E_{\rm{k}}-\Delta E_{2}$. With the set of data points for each isotope, the energy response of the CsI(Tl) can be calibrated. For $Z=1$ isotopes, the calibration function is written as \cite{hira19},

\begin{equation} \tag{3}
\begin{aligned}
L(E,Z=1,A) = a_0 + a_1 E^{(a_{2}+A)/(a_{3}+A)} \label{hydrogen}
\end{aligned}
\end{equation}
where $a_{0}$ is an offset, $a_{1}$ is a gain factor, $A$ is the mass number of $Z=1$ isotope, and $a_{2}$, $a_{3}$ are empirical non-linearity parameters.

For isotopes with $Z\ge 2$, a standard Horn's formula is used for calibration~\cite{horn92},

\begin{equation} \tag{4}
\begin{aligned} 
L(E,Z\ge2,A) = a_0 + a_1 \left( E-a_2 AZ^2{\rm log}\left( \frac{E+a_2 AZ^2}{a_2AZ^2}  \right) \right) \label{heavyion}
\end{aligned}
\end{equation}
where $a_{0}, a_{1}$ and $a_{2}$ are parameters obtained from a simultaneous fit for all the heavy isotopes. 

\begin{figure}[tp]
\centering
\includegraphics[width=0.45\textwidth]{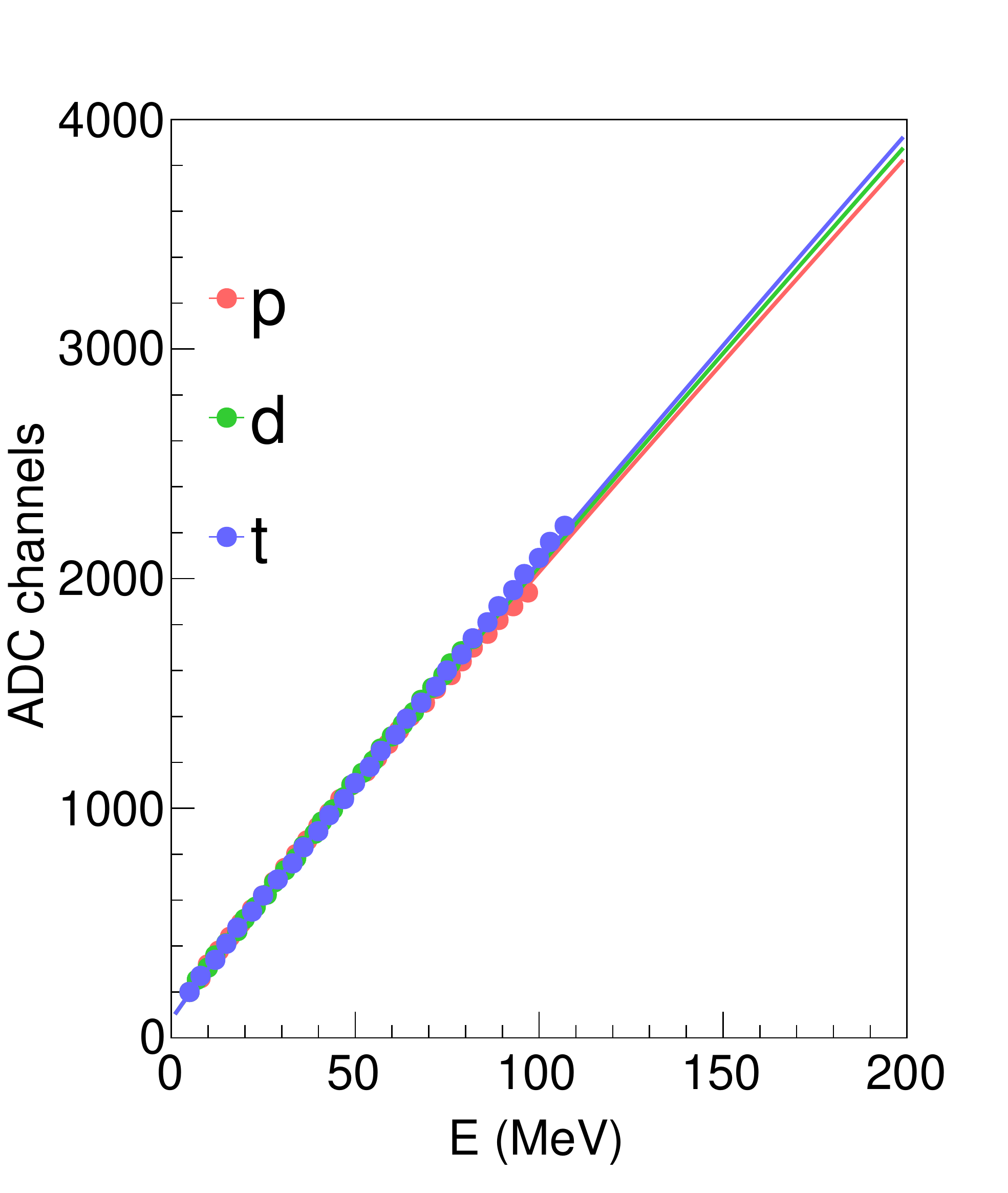} 
\caption{(Color online) Energy calibration for hydrogen isotopes proton (red points), deuteron (green points), and triton (blue points) from CsI No. 8 in telescope No. 4. The curves represent the fitting results with formula (\ref{hydrogen}).}
\label{fig:CsI_Z1}
\end{figure}

\begin{figure}[tp]
\centering
\includegraphics[width=0.45\textwidth]{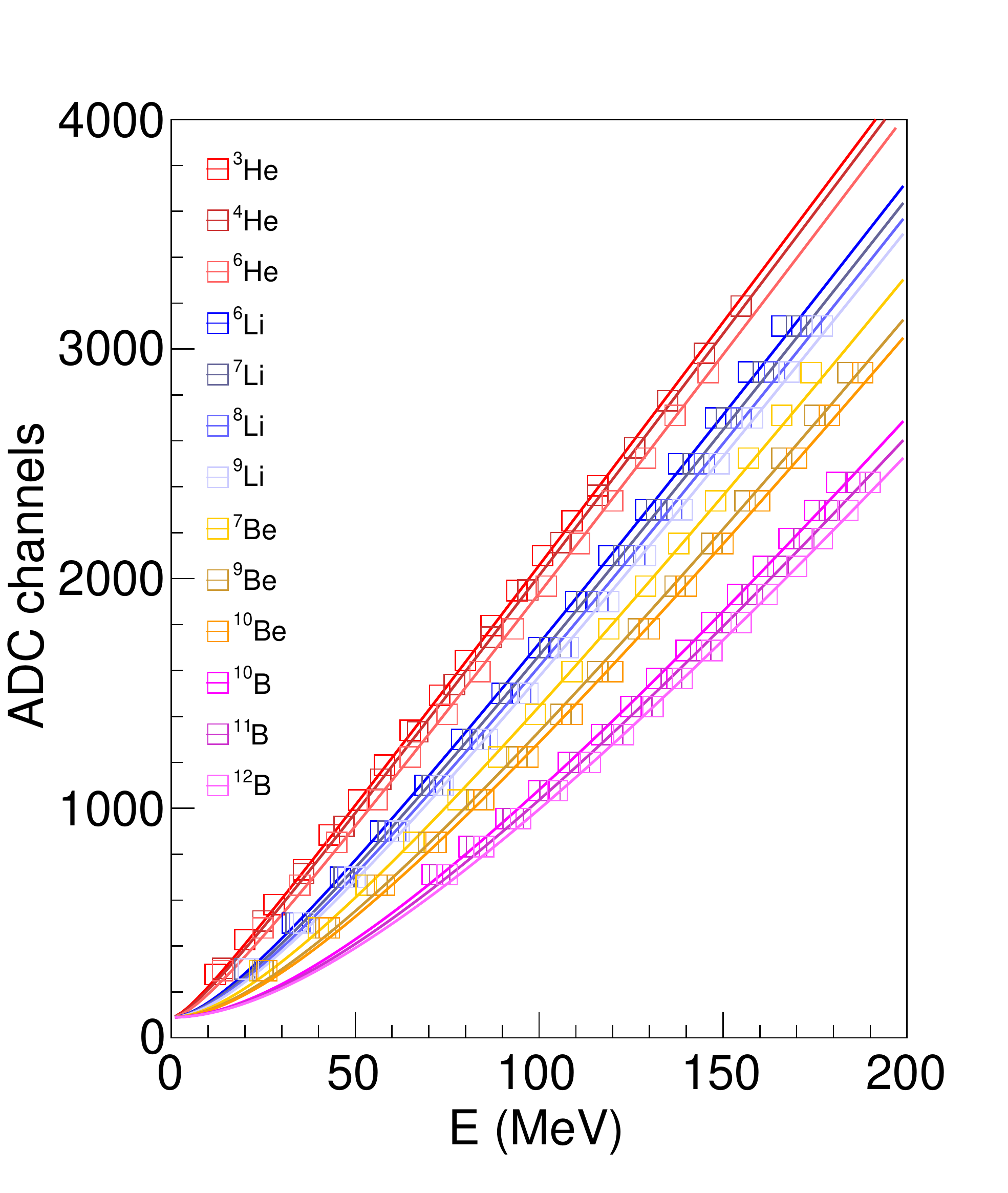} 
\caption{(Color online) Energy calibration for heavy ions $^{2,3,6}{\rm He}$, $^{6,7,8,9}{\rm Li}$, $^{7,9,10}${\rm Be} and $^{10,11,12}{\rm B}$ from CsI No. 8 in telescope No. 4. The corresponding curves represent the fitting results with formula (\ref{heavyion}).}
\label{fig:CsI_Z2}
\end{figure}

As shown in Fig. {\ref{fig:CsI_Z1}}, the energy response  for $Z=1$ isotopes exhibits insignificant non-linearity and slight difference among the isotopes in the energy range of $0-100$ MeV in accordance with the results in Ref.~\cite{hira19}. For the heavy isotopes from ${\rm He}$ to ${\rm Boron}$,  however, the non-linearity is increasingly pronounced depending on the species.  With increasing the charge of the particle with the same energy, the response of the crystal is smaller, as shown in Fig. \ref{fig:CsI_Z2}. The trends of the fitting curves are consistent with the results of HiRA10 detectors~\cite{hira19}, and a detailed discussion on the energy calibration of CsI(Tl) crystals can be found therein. 

\section{Track reconstruction for the SSD-SSD-CsI telescope}\label{sec. III}

In a CSHINE telescope, the firing position of an incident particle is determined by the crossing point of the front and the back strips of the DSSSD in the unit of pixel. The size of the pixel gives the position resolution. It is rather straightforward to recognize the  track if only one particle hits on the telescope. When multiplicity increases, however, the track finding becomes much more complicated for the following reasons, i) the signal combination of the X strips and the Y strips on the DSSSD is not unique,  ii) multi-track signals are hardly distinguishable with the charge sharing effect,  and iii) hit can be missing with certain probability.

\subsection{Constraints}\label{sec: III1}

In order to recognize all the tracks impinging on the SSD telescope, one has to introduce a set of conditions to identify the real track signals. Considering the configuration of the  CSHINE SSD telescope from top to bottom, there are four layers containing the hit information, i.e., the  SSSSD, the  DSSSD (with front- and back- side strips) and the $3 \times 3$ CsI(Tl) hodoscope. There are basically two types of conditions used for track reconstruction. One is the geometry condition marked by  $\rm{G_{O_1,O_2}}$,  the other is the energy condition marked by $\rm {E_{O_1,O_2}}$, where the subscripts ${\rm O_1}$ and ${\rm O_2}$ denote the two correlated objects that the condition is applied to.  The geometry conditions define the spatial match of one track in the SSD telescope, while the energy conditions define correspondingly the match of energy signals.   Using ``1S" to represent the SSSSD,  ``2F" (``2B") to represent the front- (back-) side strips of the DSSSD and ``3A" to represent the CsI hodoscope, repsectively, one can list all the conditions as following:
\begin{itemize}
\setlength{\itemsep}{0pt}
 \item[$\bullet$] Geometry conditions: 
     \begin{itemize}
     \setlength{\itemsep}{0pt}
       \item $\rm{G_{3A,2B}}$ (for DSSSD-CsI) 
       \item $\rm{G_{3A,2F}}$ (for DSSSD-CsI)
       \item $\rm{G_{2B,1S}}$ (for both SSSSD-DSSSD and DSSSD-CsI)
     \end{itemize}
 \item[$\bullet$] Energy conditions
     \begin{itemize}
     \setlength{\itemsep}{0pt}
       \item $\rm{E_{3A,2F}^{iso}}$ (for DSSSD-CsI)
       \item $\rm{E_{2F,1S}}$ (for DSSSD-CsI)
       \item $\rm{E_{2B,2F}}$  (within the DSSSD)
       \item $\rm{E_{2F,1S}^{iso}}$ (for SSSSD-DSSSD)
       
     \end{itemize}
\end{itemize}
The superscript ``iso" means that the condition is to match a certain isotope, see next.

\begin{table}[tp]
\renewcommand{\arraystretch}{1.}
\setlength\tabcolsep{3pt}
\caption{Geometrical map of the CSHINE telescopes. In each telescope, there are 16 channels on each silicon layer and $3 \times 3$ CsI(Tl) crystals for the array.}
\label{table:geomatch}
\centering
\small
\begin{tabular}{cccc}
\toprule
CsI No.          &``2B" Strip No.     &``2F" Strip No.    &``1S" Strip No. \\[1.5pt] 
$(N_{\rm{3A}})$   &$(N_{\rm{2B}})$     &$(N_{\rm{2F}})$    &$(N_{\rm{1S}})$ \\[1.5pt] 
\midrule
0  &$10\le N_{\rm{2B}} \le 15$  &$10\le N_{\rm{2F}} \le 15$  &\multirow{9}{*}{$\lvert N_{\rm{2B}}-N_{\rm{1S}}\rvert \le 1$}  \\
1  &$10\le N_{\rm{2B}} \le 15$  &$05\le N_{\rm{2F}} \le 10$  \\       
2  &$10\le N_{\rm{2B}} \le 15$  &$00\le N_{\rm{2F}} \le 05$  \\

3  &$05\le N_{\rm{2B}} \le 10$  &$10\le N_{\rm{2F}} \le 15$  \\
4  &$05\le N_{\rm{2B}} \le 10$  &$05\le N_{\rm{2F}} \le 10$  \\
5  &$05\le N_{\rm{2B}} \le 10$  &$00\le N_{\rm{2F}} \le 05$  \\

6  &$00\le N_{\rm{2B}} \le 05$  &$10\le N_{\rm{2F}} \le 15$  \\
7  &$00\le N_{\rm{2B}} \le 05$  &$05\le N_{\rm{2F}} \le 10$  \\
8  &$00\le N_{\rm{2B}} \le 05$  &$00\le N_{\rm{2F}} \le 05$  \\
\midrule
DSSSD-CsI  &$\rm{G_{3A,2B}}$   &$\rm{G_{3A,2F}}$   &$\rm{G_{2B,1S}}$\\
\midrule
SSSSD-DSSSD   &   &    &$\rm{G_{2B,1S}}$\\
\bottomrule
\end{tabular}
\end{table}

Table \ref{table:geomatch} lists the detailed geometrical mapping of the CSHINE SSD telescope. For the particles stopped in a given unit of the CsI(Tl) crystal marked by ${N_{\rm{3A}}}$, only certain strips of the DSSSD, marked by ${N_{\rm{2F}}}$ and ${N_{\rm{2B}}}$ for the front and back strips, respectively,  match the geometrical conditions $\rm{G_{3A,2B}}$ and $\rm {G_{3A,2F}}$. These two conditions are released for the particles stopped in the DSSSD. For example, as shown in the table, for the central CsI(Tl) unit with ${N_{\rm{3A}}=4}$, only the central part of the DSSSD with $05\le N_{\rm{2B}} \le 10$   and $05\le N_{\rm{2F}} \le 10$ satisfies the geometrical condition and delivers the signal of a same track. In addition, $\rm{G_{2B,1S}}$ is required for  geometrical  matching between  the strips of the SSSSD and the back strips of the DSSSD.  According to the configuration of the CSHINE  telescope, the strips of the SSSSD and the back strips of the DSSSD are in parallel and  numbered in the same direction. Given that the thickness of the silicon detector is very small compared to its distance to the target, one incident particle shall fire the  SSSSD and the back-side of the DSSSD with the same or neighboring strip number, as constrained by $\rm{G_{2B,1S}}$ of $\lvert N_{\rm{2B}}-N_{\rm{1S}}\rvert \le 1$.

Table \ref{table:eneconstraint} lists the energy constraints for track decoding. For an effective event with the particle stopped in the CsI(Tl) crystal, three energy conditions are used. i) The particle's energies $(E,\Delta E_{2})$ should be inside banana-shaped polygon window corresponding to  a certain isotope (discussed in sec. \ref{sec: II21}), which is denoted as $\rm{E_{3A,2F}^{iso}}$. ii) The energies detected in the front-side ($\Delta E_{\rm{2F}}$) and back-side ($\Delta E_{\rm{2B}}$) of the DSSSD should be equal in ideal case. In reality, the relative difference of $\Delta E_{\rm{2F}}$ and $\Delta E_{\rm{2B}}$ should be smaller than a given value, which is represented  by  $\rm{E_{2B,2F}}$.  iii) A rather loose condition   $\rm{E_{2F,1S}}$ can be imposed  as $\Delta  E_{\rm{2F}} > \Delta E_{\rm{1S}}$, since the energy loss in the SSSSD and the DSSSD are roughly proportional to the detector thicknesses. Similarly, for the particle stopped in the DSSSD, the condition $\rm{E_{3A,2F}^{iso}}$ can be released and only two energy conditions are left.  One is $\rm{E_{2F,1S}^{iso}}$, referring that the data point  $(E_{\rm{2F}}, \Delta E_{\rm{1S}})$  ought to be inside a certain isotope cut, the other is $\rm{E_{2B,2F}}$, referring that the signal magnitude on the front and back strips of the DSSSD ought to be approximately equal.

\begin{figure}[htp]
\centering
\includegraphics[width=0.75\textwidth]{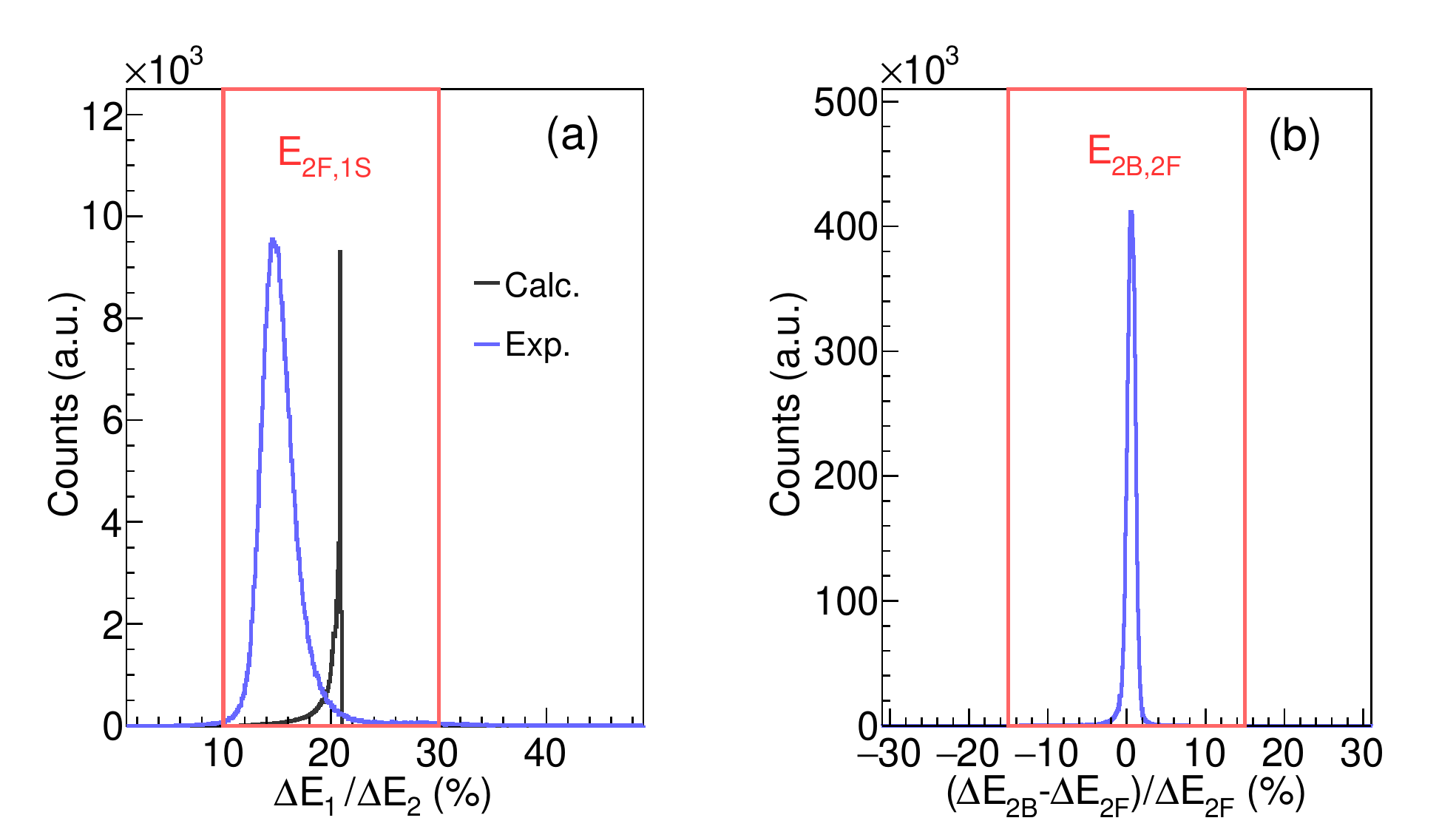} 
\caption{(Color online) Panel (a) shows the the energy loss ratios of the particles penetrating two silicon layers with the thicknesses of 110 $\rm{\mu m}$ and 526 $\rm{\mu m}$, respectively. The blue and black curves represent the experimental data and the calculated results, respectively.} Panel (b) shows the differences between the energy losses in the back- and front-side of the DSSSD extracted from the experimental data. The energy constraints $\rm{E_{2F,1S}}$ and $\rm{E_{2B,2F}}$ used for analysis are represented by the box regions, respectively.
\label{fig:enecondition}
\end{figure}

\begin{table}[t]
\renewcommand{\arraystretch}{1.2}
\setlength\tabcolsep{5pt}
\caption{Energy constraints of the CSHINE telescope No. 3. The thicknesses of the SSSSD and DSSSD are 110 $\rm{\mu m}$ and 526 $\rm{\mu m}$, respectively.}
\label{table:eneconstraint}
\centering
\small
\begin{tabular}{l|ll}
\toprule
Cases         &Constraints     & Descriptions \\[1.5pt] 
\midrule
\multirow{3}{*}{DSSSD-CsI}  &$\rm{E_{3A,2F}^{iso}}$  &$(E,\Delta E_{\rm{2F}})$ inside an isotope cut \\
\cline{2-3}
   &$\rm{E_{2F,1S}}$  &$0.10 \le \Delta E_{\rm{1S}}/ \Delta E_{\rm{2F}} \le 0.30$ $^{\rm a}$ \\ 
\cline{2-3}
   &\multirow{2}{*}{$\rm{E_{2B,2F}}$}  &\multirow{2}{*}{$\lvert (\Delta E_{\rm{2B}}-\Delta E_{\rm{2F}})/\Delta E_{\rm{2F}} \rvert \le 0.15$ $^{\rm b}$} \\
\cline{1-1}
\multirow{2}{*}{SSSSD-DSSSD} \\
   \cline{2-3}
   &$\rm{E_{2F,1S}^{iso}}$ &$(E_{\rm{2F}},\Delta E_{\rm{1S}})$ inside an isotope cut \\
\bottomrule
\end{tabular}
\leftline{\footnotesize{$^{\rm a}$ Calculated results depending on the detector thicknesses.}}
\leftline{\footnotesize{$^{\rm b}$ Extracted from the experimental data.}}
\end{table}

Fig. {\ref{fig:enecondition}} (a) shows the $\Delta E_{1}/\Delta E_{2}$ distributions, where $\Delta E_{1}$ and $\Delta E_{2}$ are energy losses of the particle penetrating two silicon detectors. The black curve is the result calculated by a numerical inversion of Ziegler's energy loss tables~\cite{ziegler}, while the blue curve represents the experimental result. It is shown that the peaks of both cases situate at 20\% and 15\%, respectively. The energy condition $\rm{E_{2F,1S}}$ is determined as $0.10 \le \Delta E_{\rm{1S}}/ \Delta E_{\rm{2F}} \le 0.30$, as shown in the box region. The relationship between $\Delta E_{\rm{2B}}$ and $\Delta E_{\rm{2F}}$ is obtained from the experimental data and is shown in Fig. {\ref{fig:enecondition}} (b). The box region represents the condition $\rm{E_{2B,2F}}$, e.g., $\lvert (\Delta E_{\rm{2B}}-\Delta E_{\rm{2F}})/\Delta E_{\rm{2F}} \rvert \le 0.15$. Descriptions of all the energy constraints are listed in Table {\ref{table:eneconstraint}}.

With the geometry and energy conditions defined above, track recognition  and event reconstruction can be conducted for the SSD-SSD-CsI telescopes. Since the events with particles stopped in the CsI hodoscope (DSSSD-CsI) and in the DSSSD (SSSSD-DSSSD) are quite different, methods of track recognition in this two cases are introduced in the following two sub-sections, respectively.

\subsection{Track reconstruction for the DSSSD-CsI} \label{sec: III2}
While track recognition for the event with track multiplicity $M_{\rm tr}=1$ in the telescope is easy,  it is much more complicated when $M_{\rm tr}$ increases. In the following text, the method for track recognition with multiple tracks ($M_{\rm tr}\ge 1$) is described, with special care of considering the charge sharing and multi-track effects. In the analysis, it is assumed that charge sharing occurs  only between two adjacent channels. For simplicity, the track recognition procedure is started  from the last detector layer (``3A") to the SSSSD  layer (``1S"), as the multiplicity always decreases from first detector layer to the last one. With all the preparation mentioned above, track reconstruction is to solve the following two questions: 

\begin{figure}[tp]
\centering
\includegraphics[width=0.52\textwidth]{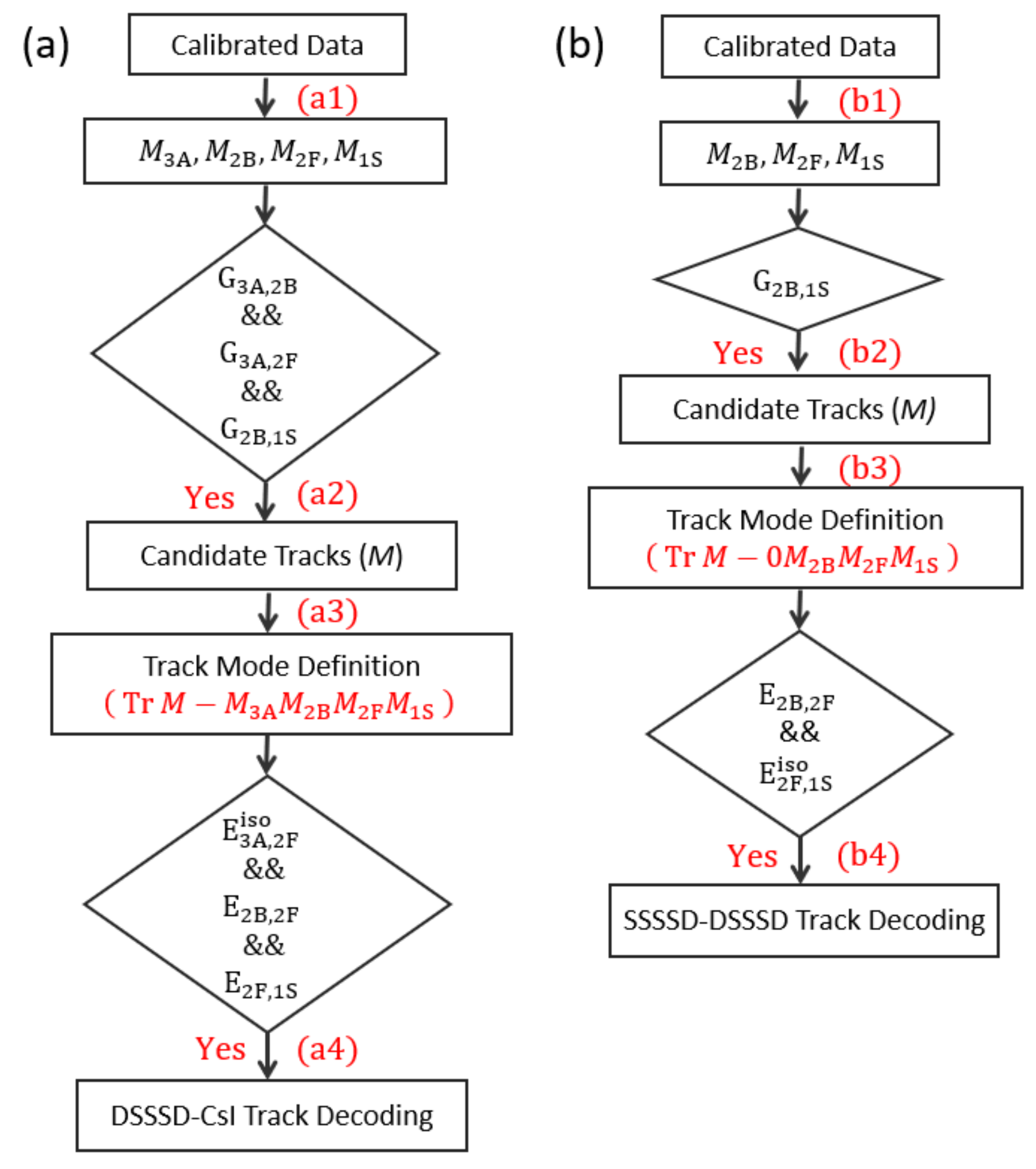} 
\caption{Flow charts of track reconstruction for (a) DSSSD-CsI and for (b) SSSSD-DSSSD.}
\label{fig:flow}
\end{figure}
\begin{itemize}
\setlength{\itemsep}{0pt}
\item[i)] When to consider charge sharing and multi-hit effects? 
\item[ii)] How to deal with charge sharing and multi-hit effects?
\end{itemize}

\subsubsection{``When": track mode definition} \label{sec: III21}
In order to find the real tracks as efficiently as possible, the track mode is classified following the  flow diagram shown in Fig. \ref{fig:flow} (a), and the steps are explained below.

Step (a1) : For each event, to extract the multiplicities of all the detector layers, $M_{\rm{3A}}, M_{\rm{2B}}, M_{\rm{2F}}$ and $M_{\rm{1S}}$, respectively. A real signal requires the amplitude to exceed a certain threshold, which is  empirically set as above the pedestal for each individual channel.

Step (a2) : To determine the number of candidate tracks ($M$) for each event with the geometrical constraints. It is an important step for track finding as only the geometrical constraints are used for determining the candidate tracks. The candidate tracks are obtained by combining all the multiplicities of the four layers to satisfy $\rm{G_{3A,2B}, G_{3A,2F}}$ and $\rm{G_{2B,1S}}$ simultaneously. The event ratios of different $M$ are listed in Appendix \ref{table:DSSSD_CsI}. As shown, track reconstruction is performed for events with $M=1,2,3,4$ and 6 (${M=5}$ is omitted as the percentage is much smaller). 

Step (a3) : To define the track modes with $M$ as well as the multiplicities, denoted as $``{\rm{Tr}}M-M_{\rm{3A}}M_{\rm{2B}}M_{\rm{2F}}M_{\rm{1S}}"$. For $M \ge 1$, one has $1 \le M_{\rm{3A}} \le M$, $1 \le M_{\rm{2B}} \le M$, $1 \le M_{\rm{2F}} \le M$ and $1 \le M_{\rm{1S}} \le M$, and there are  maximally $M^{4}$ different track modes in total. However, it is found that the number of the effective modes are far less than $M^{4}$. In track decoding, we selected the top modes as shown in Appendix Table {\ref{table:DSSSD_CsI}}.

With the track modes well-defined, one can decode every mode according to the characteristics of the data and identify the charge sharing and multi-hit effects.

\subsubsection{``How": track mode decoding} \label{sec: III22}
For DSSSD-CsI, a track is well-defined when both $\rm{E_{3A,2F}^{iso}}$ and $\rm{E_{2B,2F}}$ are true, while $\rm{E_{2F,1S}}$ is only used to cross check the  energy loss of the particle in the first silicon layer. The cases of track decoding are listed as follows:

i) Consider charge sharing effect. In the track decoding, it is assumed that charge sharing only happens between two adjacent channels in the same layer. For $M=2$, it is easy to find the track with charge sharing. We take one Tr2-1211 event for an example, as shown in Table \ref{table:typicalmodes}. This is an event with $M_{\rm{2B}}=2$ while $M_{\rm{3A}}=M_{\rm{2F}}=M_{\rm{1S}}=1$. The number of candidate track is $M=2$. It is shown that the sum of the energy losses in the two neighboring strips 8 and 9  on ``2B" layer is close to the energy loss in the single strip 13 on ``2F" layer, which satisfies the energy condition $\rm{E_{2B,2F}}$. If $\rm{E_{2F,3A}^{iso}}$ is also true, then one real track is finally determined. For even larger M, we have to find out two tracks with adjacent strips so that we can consider charge sharing effect. The example is also shown in Table \ref{table:typicalmodes} by a Tr4-1212 event. The number of track candidates is $M=4$, and two firing strips are found on both 2B and 1S layers. However, charge sharing in ``2B" layer can be identified by comparing track candidate  No. 2 and No. 3 (or No. 1 and No. 4), indicating that there is only one valid track in this mode. The ratios of charge sharing for the track modes are listed in Appendix \ref{table:DSSSD_CsI}.

\begin{table}[!hp]
\renewcommand{\arraystretch}{0.8}
\setlength\tabcolsep{10pt}
\caption{ Data characteristics of some typical track modes. }
\label{table:typicalmodes}
\centering
\scriptsize
\begin{tabular}{ccccccccc}
\toprule
\\
\multicolumn{9}{c}{Charge sharing: $\rm{Tr2-1211}$} \\
\midrule
Track No. &$N_{\rm{3A}}$  &$N_{\rm{2B}}$  &$N_{\rm{2F}}$  &$N_{\rm{1S}}$  &$E_{\rm{3A}}$ &$\Delta E_{\rm{2B}}$ &$\Delta  E_{\rm{2F}}$ &$\Delta  E_{\rm{1S}}$  \\
&&&& &$\rm{(ch)}$ &$\rm{(MeV)}$  &$\rm{(MeV)}$  &$\rm{(MeV)}$      \\
\midrule
1  &3  &\textcolor{red}{8}  &13  &9  &1273  &\textcolor{red}{58.15}  &98.12  &13.40\\
2  &3  &\textcolor{red}{9}  &13  &9  &1273  &\textcolor{red}{36.58}  &98.12  &13.40\\
\\

\multicolumn{9}{c}{Charge sharing: $\rm{Tr4-1212}$} \\
\midrule
Track No. &$N_{\rm{3A}}$  &$N_{\rm{2B}}$  &$N_{\rm{2F}}$  &$N_{\rm{1S}}$  &$E_{\rm{3A}}$ &$\Delta E_{\rm{2B}}$ &$\Delta  E_{\rm{2F}}$ &$\Delta  E_{\rm{1S}}$  \\
&&&& &$\rm{(ch)}$ &$\rm{(MeV)}$  &$\rm{(MeV)}$  &$\rm{(MeV)}$      \\
\midrule
1  &6  &2  &15  &2  &1014  &19.62  &26.78  &4.21  \\
2  &6  &\textcolor{red}{2}  &15  &\textcolor{red}{3}  &1014  &\textcolor{red}{19.62}  &26.78  &\textcolor{red}{1.63} \\
3  &6  &\textcolor{red}{3}  &15  &\textcolor{red}{2}  &1014  &\textcolor{red}{7.18}  &26.78  &\textcolor{red}{4.21}  \\
4  &6  &3  &15  &3  &1014  &7.18  &26.78  &1.63 \\
\\

\multicolumn{9}{c}{Multi-hit: $\rm{Tr2-2212}$} \\
\midrule
Track No. &$N_{\rm{3A}}$  &$N_{\rm{2B}}$  &$N_{\rm{2F}}$  &$N_{\rm{1S}}$  &$E_{\rm{3A}}$ &$\Delta E_{\rm{2B}}$ &$\Delta  E_{\rm{2F}}$ &$\Delta  E_{\rm{1S}}$  \\
&&&& &$\rm{(ch)}$ &$\rm{(MeV)}$  &$\rm{(MeV)}$  &$\rm{(MeV)}$      \\
\midrule
1  &5  &6  &\textcolor{red}{2}  &6  &1163  &\textcolor{red}{12.01}  &38.45  &1.55 \\
2  &8  &3  &\textcolor{red}{2}  &3  &1890  &\textcolor{red}{26.56}  &38.45  &3.49 \\
\\

\multicolumn{9}{c}{Multi-hit: $\rm{Tr4-2122}$} \\
\midrule
Track No. &$N_{\rm{3A}}$  &$N_{\rm{2B}}$  &$N_{\rm{2F}}$  &$N_{\rm{1S}}$  &$E_{\rm{3A}}$ &$\Delta E_{\rm{2B}}$ &$\Delta  E_{\rm{2F}}$ &$\Delta  E_{\rm{1S}}$  \\
&&&& &$\rm{(ch)}$ &$\rm{(MeV)}$  &$\rm{(MeV)}$  &$\rm{(MeV)}$      \\
\midrule
1  &7  &3  &\textcolor{red}{9}  &3  &1670  &16.12  &\textcolor{red}{9.98}  &1.91  \\
2  &7  &3  &9  &4  &1670  &16.12  &9.98  &0.90  \\
3  &8  &3  &2  &3  &3444  &16.12  &6.01  &1.91  \\
4  &8  &3  &\textcolor{red}{2}  &4  &3444  &16.12  &\textcolor{red}{6.01}  &0.90  \\
\\

\multicolumn{9}{c}{No charge sharing or multi-hit: ${\rm Tr4-2222^I}$} \\
\midrule
Track No. &$N_{\rm{3A}}$  &$N_{\rm{2B}}$  &$N_{\rm{2F}}$  &$N_{\rm{1S}}$  &$E_{\rm{3A}}$ &$\Delta E_{\rm{2B}}$ &$\Delta  E_{\rm{2F}}$ &$\Delta  E_{\rm{1S}}$  \\
&&&& &$\rm{(ch)}$ &$\rm{(MeV)}$  &$\rm{(MeV)}$  &$\rm{(MeV)}$      \\
\midrule
1  &5  &8  &0  &8  &822   &\textcolor{red}{14.83}  &\textcolor{red}{14.77}  &1.90\\
2  &5  &8  &4  &8  &822   &14.83  &37.52  &1.90\\
3  &8  &3  &0  &3  &1012  &37.57  &14.77  &5.48\\
4  &8  &3  &4  &3  &1012  &\textcolor{red}{37.57}  &\textcolor{red}{37.52}  &5.48\\
\\

\multicolumn{9}{c}{No charge sharing or multi-hit: ${\rm Tr4-2222^{II}}$} \\
\midrule
Track No. &$N_{\rm{3A}}$  &$N_{\rm{2B}}$  &$N_{\rm{2F}}$  &$N_{\rm{1S}}$  &$E_{\rm{3A}}$ &$\Delta E_{\rm{2B}}$ &$\Delta  E_{\rm{2F}}$ &$\Delta  E_{\rm{1S}}$  \\
&&&& &$\rm{(ch)}$ &$\rm{(MeV)}$  &$\rm{(MeV)}$  &$\rm{(MeV)}$      \\
\midrule
1  &2  &12  &0  &12  &724   &\textcolor{red}{16.34}  &\textcolor{red}{16.93}  &1.95\\
2  &2  &12  &1  &12  &724   &\textcolor{red}{16.34}  &\textcolor{red}{16.31}  &1.95\\
3  &5  &9   &0  &9   &641   &\textcolor{red}{17.04}  &\textcolor{red}{16.93}  &2.00\\
4  &5  &9   &1  &9   &641   &\textcolor{red}{17.04}  &\textcolor{red}{16.31}  &2.00\\
\\

\multicolumn{9}{c}{Cross check $\Delta E_{\rm{1S}}$: $\rm{Tr2-2222}$} \\
\midrule
Track No. &$N_{\rm{3A}}$  &$N_{\rm{2B}}$  &$N_{\rm{2F}}$  &$N_{\rm{1S}}$  &$E_{\rm{3A}}$ &$\Delta E_{\rm{2B}}$ &$\Delta  E_{\rm{2F}}$ &$\Delta  E_{\rm{1S}}$  \\
&&&& &$\rm{(ch)}$ &$\rm{(MeV)}$  &$\rm{(MeV)}$  &$\rm{(MeV)}$      \\
\midrule
1  &2  &13  &3   &13  &1573  &9.99   &9.97   &\textcolor{red}{1.44} \\
2  &3  &5   &13  &5   &1264  &12.84  &12.80  &\textcolor{red}{19.34} \\
\\
\bottomrule
\end{tabular}
\end{table}

ii) Consider multi-hit effect. When ``multi-hit" is mentioned, it refers that there are two (or more) particles hitting the same strip. In the analysis, events with two or more particles hitting the same CsI(Tl) crystal are discarded as we can not get the real deposited energy in the crystal in this case. In Table \ref{table:typicalmodes}, e.g., Tr2-2212 is an event with two candidate tracks which are all valid, with the two particles hitting the same strip on the ``2F" layer. In such case $E_{\rm{2B}}$ is used instead of $E_{\rm{2F}}$ for particle identification when double hits happen on the front-side of the DSSSD. For larger $M$, we have to try every two tracks to find out the positions of the hits. To see Tr4-2122 in Table \ref{table:typicalmodes} for another example, track No. 1 and No. 4 (or track No. 2 and No. 3) are valid tracks with two particles hitting the same strip on the back-side ``2B" of the DSSSD. The ratios of multi-hit events for all the track modes are listed in Appendix \ref{table:DSSSD_CsI}.

iii) No charge sharing or double-hit effects. In most cases, instead of considering charge sharing and double-hit effects, we try to find out the real tracks with $\rm{E_{3A,2F}^{iso}}$ and $\rm{E_{2B,2F}}$ directly. The number of valid tracks is determined by the smallest multiplicity among $M_{\rm{3A}}, M_{\rm{2B}}$ and $M_{\rm{2F}}$. As expected, there will be one real track for Tr2-1212 if no charge sharing or multi hits happen. For larger $M$, we can use $\rm{E_{2B,2F}}$ to find out the valid tracks. For example, for the event ${\rm Tr4-2222^I}$ in Table \ref{table:typicalmodes}, it is clear that track No. 1 and No. 4 are two valid tracks. However, it is very difficult to find out the real tracks if two particles with close energies hit the DSSSD, as shown the second ${\rm Tr4-2222^{II}}$ in Table \ref{table:typicalmodes}. In this case, the events are currently discarded as the statistics of this track mode is rather low.

iv) Cross check energy loss $\Delta E_{\rm{1S}}$ in the ``1S" layer. From i) to iii), we can determine the valid tracks, but we have to cross check $\Delta E_{\rm{1S}}$ as well to finalize the track reconstruction for DSSSD-CsI. See Tr2-2222 in Table \ref{table:typicalmodes} for example, track No. 1 is a valid track and satisfied with $\rm{E_{2F,1S}}$ as listed in Table \ref{table:eneconstraint}. Track No. 2 is also a valid track since it is satisfied with both $\rm{E_{3A,2F}^{iso}}$ and $\rm{E_{2B,2F}}$, but $\rm{E_{2F,1S}}$ is not true due to the large energy loss in the ``1S" layer, which possibly due to a heavy fragment stopped in the same ``1S" strip. In this case, as long as a valid track is identified, $\Delta E_{1S}$ can be recalculated by the following procedures: 1) the charge (Z) and mass (A) of the particle can be obtained from $\rm{E_{3A,2F}^{iso}}$; 2) the energy of the particle impinging the DSSSD $E_{2}^{\rm{imp}}$ can be deduced by the numerical inversion of Ziegler's energy loss tables~\cite{ziegler}; 3) the energy of the particle impinging the SSSSD $E_{1}^{\rm{imp}}$ can also be deduced similarly; 4) the particle's energy loss in the SSSSD is $E_{1}^{\rm{imp}}-E_{2}^{\rm{imp}}$. Finally, the total kinetic energy of the impinging particle is $E_{\rm{tot}} = \Delta E_{1}+\Delta E_{2} + E$.

\begin{figure*}[t] 
  \centering
  \includegraphics[width=0.96\textwidth,height=0.28\textheight]{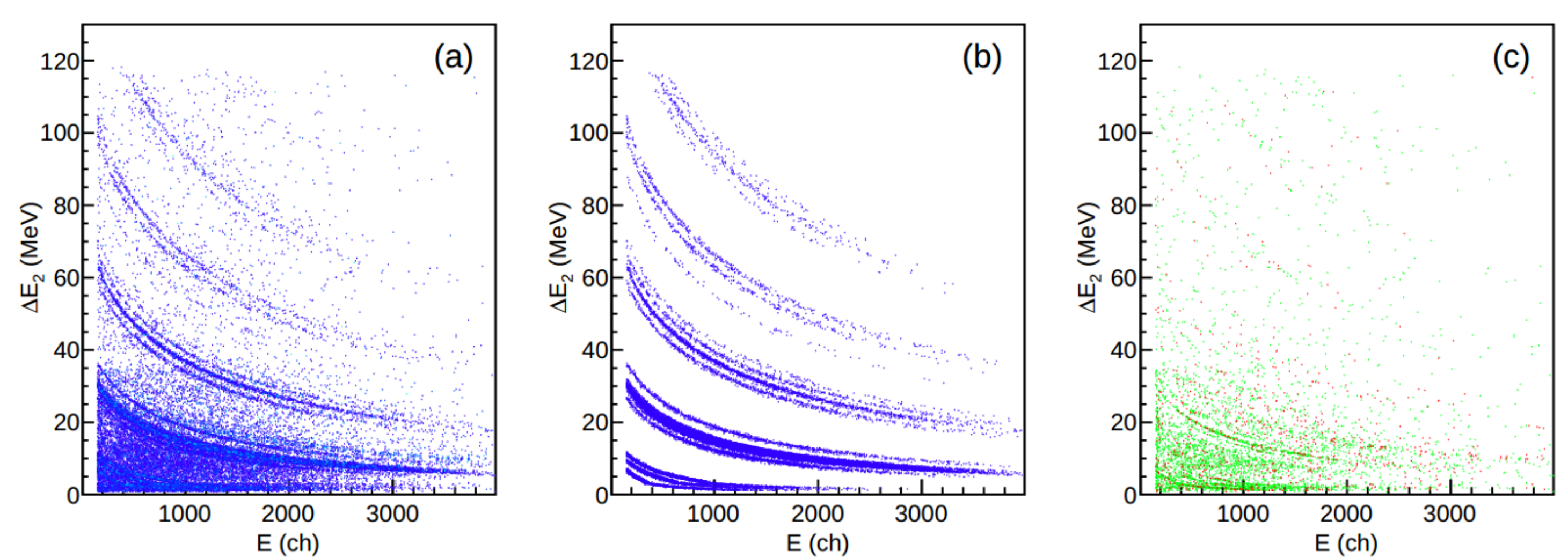} 
  \caption{Results of track reconstruction for DSSSD-CsI of telescope No. 3. Panel (a) shows the results for the tracks with $M=1, 2, 3, 4$, and $6$. Panel (b) represents the results after track decoding, and panel (c) shows the lost events.}
  \label{fig:Si_CsI_Track}
\end{figure*}

Based on the above procedures, we can decode the tracks effectively. As shown in Appendix \ref{table:DSSSD_CsI}, the total efficiency of the track decoding for DSSSD-CsI is 88.59 $\%$, and the ratios of charge sharing and multi-hit are 6.60 $\%$ and 1.13 $\%$, respectively. Comparing with the test results  with an $\alpha$ source, where the ratio of charge sharing is found to be less than 1 $\%$, the charge sharing effect is much more obvious in the beam experiment, indicating this effect must be included in data analysis. Fig. \ref{fig:Si_CsI_Track} (a) and (b) present the $\Delta E_2-E$ scattering plots of telescope No. 3 before and after the track decoding, respectively. For the DSSSD-CsI, the $Z>5$ isotopes are discarded due to their low statistics and the limited PID capability.
For a comparison, Fig. \ref{fig:Si_CsI_Track} (c) presents the results of the leftover events. It is evident that the track decoding efficiency is high.

\subsection{Track reconstruction for the SSSSD-DSSSD}\label{sec: III3}
As shown in Fig. \ref{fig:flow} (b), the procedures of track reconstruction for SSSSD-DSSSD is similar to DSSSD-CsI, but they are different in three aspects:

i) Only $\rm{G_{2B,1S}}$ is used to determine candidate tracks ($M$), because the CsI(Tl) hodoscope is not fired.

ii) The track modes is defined as $``{\rm{Tr}}M-0M_{\rm{2B}}M_{\rm{2F}}M_{\rm{1S}}"$ for the same reason. 

iii) Both $\rm{E_{2B,2F}}$ and $\rm{E_{2F,1S}^{iso}}$ are used for track decoding.

\begin{figure*}[t] 
  \centering
  \includegraphics[width=0.96\textwidth,height=0.28\textheight]{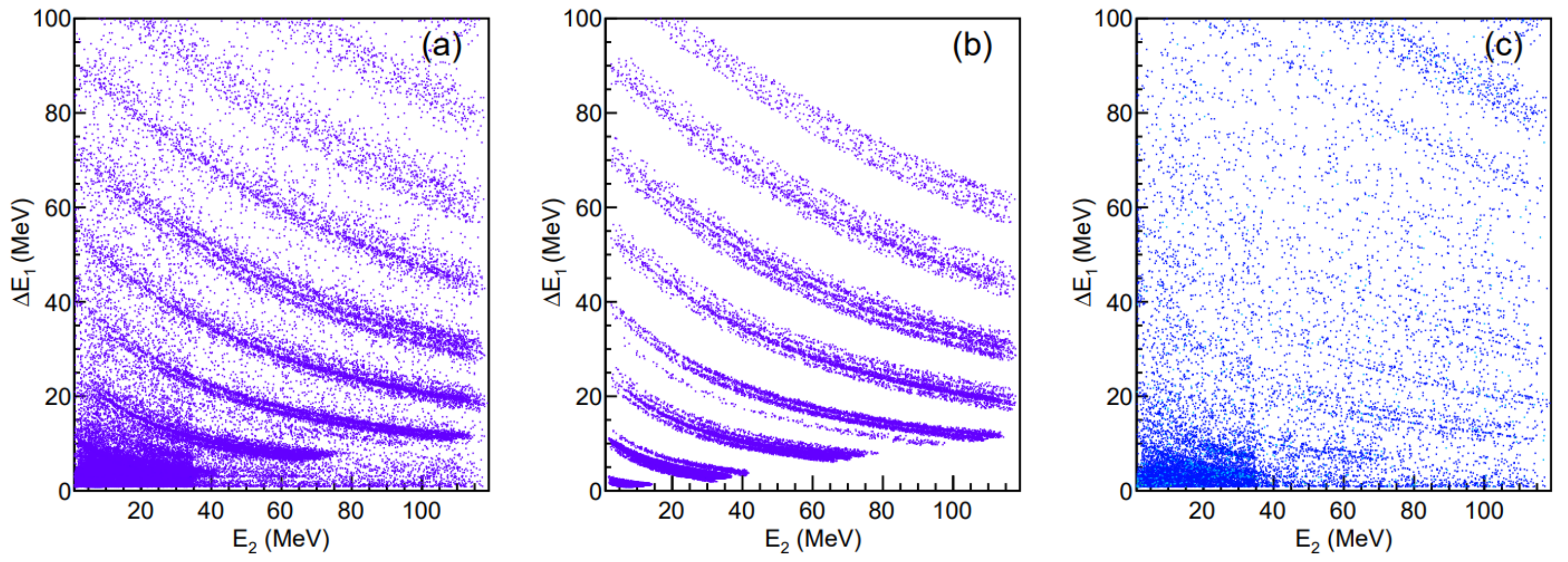} 
  \caption{Results of track reconstruction for SSSSD-DSSSD of telescope No. 3. Panel (a) shows the results for the tracks with $M=1, 2, 3, 4$, and $6$. Panel (b) represents the results after track decoding, and panel (c) shows the lost events.}
  \label{fig:Si_Si_Track}
\end{figure*}

The maximum number of track modes is $M^{3}$,  much less than those  for DSSSD-CsI. Appendix \ref{table:SSSSD_DSSSD} shows the information of track decoding for SSSSD-DSSSD. The track decoding efficiency is 79.35 $\%$, which is smaller than that for DSSSD-CsI. There are mainly two reasons for the lower efficiency, i) many heavy fragments are stopped in the first SSSSD layer and increases the multiplicity in the layer of SSSSD ($M_{\rm{1S}}$) and ii) the tracks penetrating the two SSDs but hitting on the materials in the gaps (about 1 mm wide) between the CsI crystals are mostly discarded. For the latter case, the particle may penetrates the DSSSD and escapes without energy response in the CsI crystals. These events are included in the $\Delta E_{1}-E_{2}$ plot, but are discarded in track decoding particularly in the area of low $\Delta E_1$, which may reduce the efficiency. The ratios of event with charge sharing and double-hit are 11.54 $\%$ and 0.23 $\%$, respectively. Fig. \ref{fig:Si_Si_Track} shows the results of track reconstruction for SSSSD-DSSSD of telescope No. 3. In Fig. \ref{fig:Si_Si_Track} (b), the isotopes with $Z>8$ are discarded due to their low statistics and the limited PID capability of the telescope. Comparing Fig. \ref{fig:Si_Si_Track} (b) and Fig. \ref{fig:Si_Si_Track} (c), it can be found that most of the tracks are effectively reconstructed with this method.

\section{Results}\label{sec. V}
\begin{figure*}[htbp] 
  \centering 
  \includegraphics[width=0.96\textwidth,height=0.45\textheight]{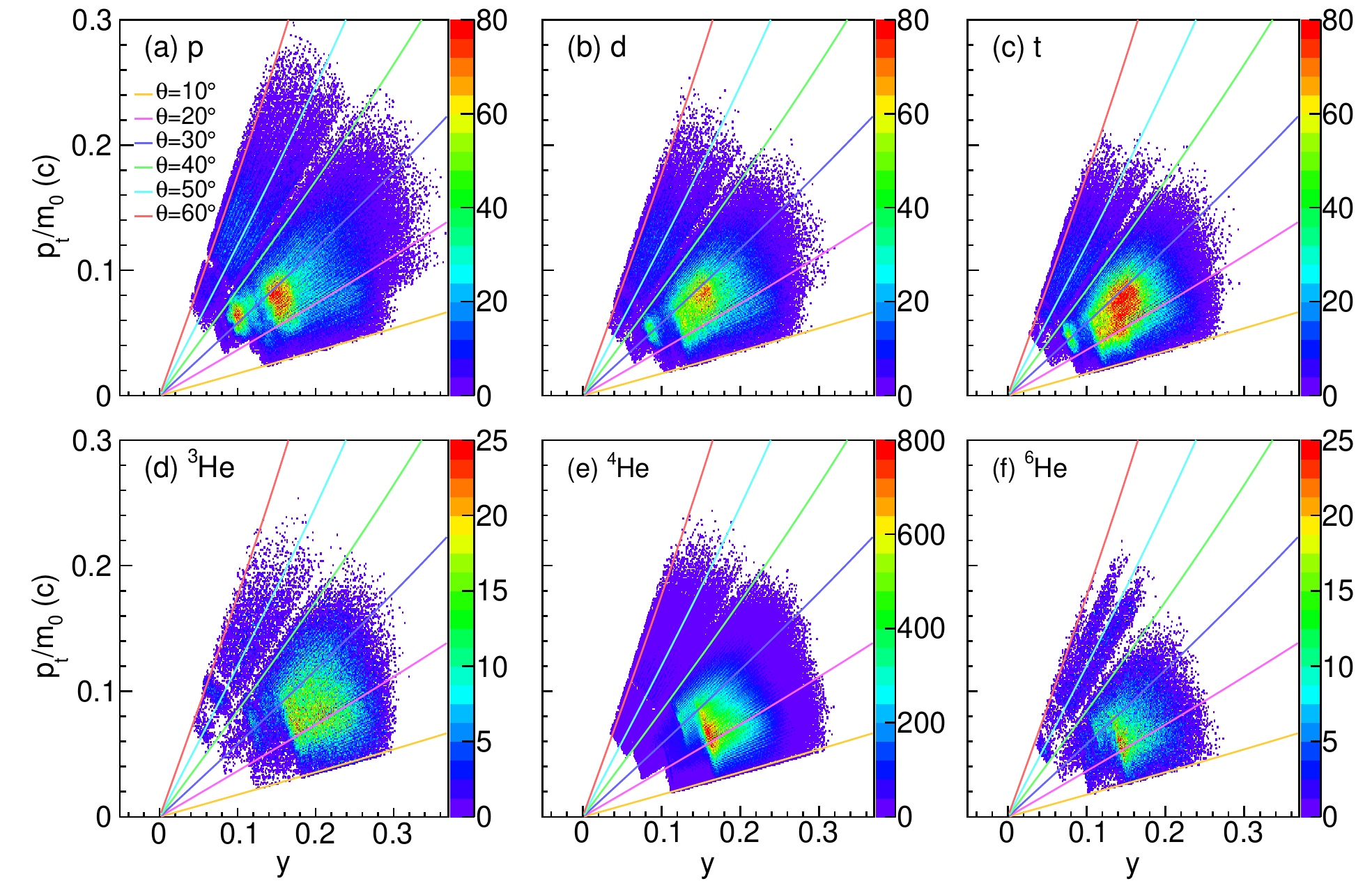} 
  \caption{(Color online) Phase space distribution of the LCPs measured by the CSHINE telescope array used in the $\rm{^{86}Kr + ^{208}Pb}$ reactions at 25 MeV/u~\cite{gfh21}. From (a) to (f) shows the results of p, d, t, $\rm{^{3}He}$, $\rm{^{4}He}$ and $\rm{^{6}He}$, respectively.}
  \label{fig:phasespace}
\end{figure*}

When the track recognition is done, one can obtain the phase space distributions for each isotopes. Fig. \ref{fig:phasespace} presents the raw phase space distribution of LCPs with $Z\le2$ for the CSHINE telescopes in the experiment with $\rm{^{86}Kr + ^{208}Pb}$  at 25 MeV/u ~\cite{gfh21}. Here the geometric efficiency arising from the incomplete azimuthal coverage is not corrected. The rapidity, the transverse momentum and the mass of the particle are denoted as $y$, $p_{t}$ and $m_0$. The curves are the results of $p_{t}/m_{0}$ as a function of $\theta$ and $y$ as , 
\begin{equation} \tag{5}
\begin{aligned}
\frac{p_{t}}{m_0} = \frac{\tanh{y}\cdot\tan\theta}{\sqrt{1-\left( \tanh{y}/\cos\theta \right)^2}}
\end{aligned}
\end{equation}
It shows that the telescopes can measure light charged particles in a wide $\theta$ angle range from $10^{\circ}$ to $60^{\circ}$.  For the current beam energies,  the emission of light charged particles in the angular range is contributed mainly by the intermediate velocity source and carries the information of the transport of the isospin degree of freedom \cite{zy17}. Therefore CSHINE with the full configuration of SSD telescopes provides opportunities for physics studies  such as the constraint of nuclear equation of state via the isospin composition of the particles as well as the HBT chronology for different isotopes  in heavy ion reactions at Fermi energies \cite{wyjplb21,dxy21}. 

\section{Summary}\label{sec. VI}
In this paper,  a novel method of track recognition for the SSD-SSD-CsI telescope is described. The procedure includes detector calibration and track reconstruction. The silicon strip detectors are calibrated using pulse and $\alpha$ source, while the CsI(Tl) crystals are calibrated using the $\Delta E_2-E$ scattering plot after PID is achieved. A novel practical method is invented for track reconstruction by a detailed analysis on the track mode, which are defined by the number of candidate tracks and the hit multiplicities in the four layers of the telescope. Through track decoding algorithm,  one can discriminate the charge sharing and multi-hit effects quantitatively. High tracking efficiency of 88.59\% and 79.35\%  is achieved in SSSSD-DSSSD and DSSSD-CsI combinations, respectively. Phase space distribution for the LCPs with $Z\le2$ are obtained, demonstrating that  CSHINE telescopes  can measure the LCPs  in the  range of $\theta=\rm{10^{\circ}}$ to $\rm{60^{\circ}}$  in laboratory and provide opportunities for future physical studies,  such as the constraint of nuclear equation of state as well as HBT chronology in heavy ion reactions at Fermi energies.

\section*{Acknowledgments}
This work is supported by the National Natural Science Foundation of China under Grants Nos. 11875174, 11961131010, 11961141004, and 11890712, the Ministry of Science and Technology under Grant No. 2020YFE0202001, the Initiative Scientific Research Program of Tsinghua University, and by the Heavy Ion Research Facility in Lanzhou (HIRFL). The authors acknowledge the gas detector group and the RIBLL group at the Institute of Modern Physics, Chinese Academy of Sciences, for offering local help in experiment and the machine staff for delivering the beam.

\appendix
\renewcommand{\thetable}{A\arabic{table}} 
\setcounter{table}{0}
\section{Modes for DSSSD-CsI of telescope No.3.}\label{sec:appendix_A}

\begin{table*}[]
\footnotesize
\renewcommand{\arraystretch}{0.7}
\setlength\tabcolsep{0.2mm}
\caption{ Modes included in track reconstruction for DSSSD-CsI of telescope No. 3.}
\label{table:DSSSD_CsI}
\centering

\begin{tabular}{c|c|c|c|c|c|c|c|c|c}
\toprule
\multirow{2}{*}{$M$} &Ratio &Decoded &Sharing &Multi-hit &\multicolumn{5}{c}{Track reconstruction}  \\[1.5pt] 
\cline{6-10} 
&$(\%)$ &$(\%)$ &$(\%)$ &$(\%)$ &modes &ratio(\%)  &decoded(\%) &sharing (\%) &multi-hit(\%) \\[1.5pt] 
\midrule
$M \ge 1$ &100  &--- &--- &--- &--- &--- &--- &--- &---\\
\midrule
\\
\midrule
$M=1$     &69.10 &93.92  &--- &--- &$\rm{Tr1-1111}$  &100 &93.92 &--- &---\\
\midrule
\multirow{8}{*}{$M=2$} &\multirow{8}{*}{17.73} &\multirow{8}{*}{87.98} &\multirow{8}{*}{29.98} &\multirow{8}{*}{4.86}  &$\rm{Tr2-1112}$  &5.07   &82.83  &18.48  &---  \\
   &&&& &$\rm{Tr2-1121}$  &54.09  &84.80  &40.23  &---\\
   &&&& &$\rm{Tr2-1211}$  &11.35  &86.85  &64.18  &--- \\
   &&&& &$\rm{Tr2-1212}$  &7.23   &86.09  &0.02   &---\\
   &&&& &$\rm{Tr2-2111}$  &0.46   &74.72  &---    &---\\
   &&&& &$\rm{Tr2-2121}$  &7.10   &96.97  &---    &60.22\\
   &&&& &$\rm{Tr2-2212}$  &0.92   &95.75  &---    &63.61\\
   &&&& &$\rm{Tr2-2222}$  &13.79  &99.51  &---    &---\\
\midrule
\multirow{14}{*}{$M=3$} &\multirow{14}{*}{2.35} &\multirow{14}{*}{84.13} &\multirow{14}{*}{17.29} &\multirow{14}{*}{7.90}  &$\rm{Tr3-1113}$          &0.17   &88.46   &---    &---\\
   &&&& &$\rm{Tr3-1131}$  &17.06  &78.12   &38.70  &---\\
   &&&& &$\rm{Tr3-1212}$  &9.73   &84.68   &9.41   &---\\
   &&&& &$\rm{Tr3-1213}$  &1.09   &81.44   &---    &---\\
   &&&& &$\rm{Tr3-1312}$  &2.76   &88.65   &31.21  &---\\
   &&&& &$\rm{Tr3-2121}$  &7.59   &87.34   &---    &49.00\\
   &&&& &$\rm{Tr3-2131}$  &8.57   &71.93   &---    &48.74\\
   &&&& &$\rm{Tr3-2221}$  &2.41   &91.33   &---    &---\\
   &&&& &$\rm{Tr3-2222}$  &23.00  &99.29   &---    &---\\
   &&&& &$\rm{Tr3-2223}$  &1.86   &98.24   &---    &---\\
   &&&& &$\rm{Tr3-2232}$  &13.27  &99.01   &40.79  &---\\
   &&&& &$\rm{Tr3-2322}$  &4.65   &97.61   &75.14  &---\\
   &&&& &$\rm{Tr3-3222}$  &1.82   &81.36   &---    &---\\
   &&&& &$\rm{Tr3-3333}$  &0.26   &100     &---    &---\\
\midrule
\multirow{8}{*}{$M=4$} &\multirow{8}{*}{6.05} &\multirow{8}{*}{80.28} &\multirow{8}{*}{14.52} &\multirow{8}{*}{1.10} &$\rm{Tr4-1122}$         &4.40   &54.93   &21.38   &---\\
   &&&& &$\rm{Tr4-1212}$ &9.91   &88.37   &88.36   &---\\
   &&&& &$\rm{Tr4-1221}$ &13.35  &79.81   &36.12   &---\\
   &&&& &$\rm{Tr4-1222}$ &22.08  &72.68   &---     &---\\
   &&&& &$\rm{Tr4-2122}$ &1.38   &89.67   &---     &79.89\\
   &&&& &$\rm{Tr4-2221}$ &8.48   &95.15   &---     &---\\
   &&&& &$\rm{Tr4-2222}$ &33.74  &97.95   &---     &---\\
   &&&& &$\rm{Tr4-3333}$ &0.05   &94.74   &---     &---\\
\midrule
$M=5$ &\sout{0.36} &--- &--- &--- &--- &--- &--- &--- &---\\
\midrule
\multirow{14}{*}{$M=6$} &\multirow{14}{*}{1.77} &\multirow{14}{*}{71.07} &\multirow{14}{*}{---} &\multirow{14}{*}{0.74}  &$\rm{Tr6-1123}$         &0.06  &14.29   &---   &---\\
   &&&& &$\rm{Tr6-1132}$ &1.20  &35.51   &---   &---\\
   &&&& &$\rm{Tr6-1222}$ &14.68 &64.26   &---   &---\\
   &&&& &$\rm{Tr6-1223}$ &1.49  &71.35   &---   &---\\
   &&&& &$\rm{Tr6-1231}$ &3.42  &64.47   &---   &---\\
   &&&& &$\rm{Tr6-1232}$ &6.10  &66.52   &---   &---\\
   &&&& &$\rm{Tr6-1322}$ &3.81  &67.43   &---   &---\\
   &&&& &$\rm{Tr6-2122}$ &0.93  &99.07   &---   &79.44\\
   &&&& &$\rm{Tr6-2221}$ &5.63  &92.44   &---   &---\\
   &&&& &$\rm{Tr6-2222}$ &26.55 &98.04   &---   &---\\
   &&&& &$\rm{Tr6-2223}$ &2.22  &94.14   &---   &---\\
   &&&& &$\rm{Tr6-2232}$ &11.81 &92.86   &---   &---\\
   &&&& &$\rm{Tr6-2322}$ &6.11  &93.60   &---   &---\\
   &&&& &$\rm{Tr6-3333}$ &0.41  &91.49   &---   &---\\
\midrule
Total &97.00 &88.59 &6.60 &1.13 &\multicolumn{5}{c}{---} \\
\bottomrule
\end{tabular}
\end{table*}

\section{Modes for SSSSD-DSSSD of telescope No.3.} \label{sec:appendix_B}
\renewcommand{\thetable}{B\arabic{table}} 
\setcounter{table}{0}
\begin{table*}[]
\footnotesize
\renewcommand{\arraystretch}{0.7}
\setlength\tabcolsep{0.2mm}
\caption{Modes included in track reconstruction for SSSSD-DSSSD of telescope No. 3.}
\label{table:SSSSD_DSSSD}
\centering

\begin{tabular}{c|c|c|c|c|c|c|c|c|c}
\toprule
\multirow{2}{*}{$M$} &Ratio &Decoded &Sharing &Multi-hit &\multicolumn{5}{c}{Track reconstruction}  \\[1.5pt] 
\cline{6-10} 
&$(\%)$ &$(\%)$ &$(\%)$ &$(\%)$ &modes &ratio(\%)  &decoded(\%) &sharing (\%) &multi-hit(\%) \\[1.5pt] 

\midrule
$M \ge 1$ &100  &--- &--- &--- &--- &--- &--- &--- &---\\
\midrule
~\\
\midrule
$M=1$     &76.28 &80.63  &--- &--- &$\rm{Tr1-0111}$  &100 &80.63 &--- &---\\
\midrule
\multirow{4}{*}{$M=2$} &\multirow{4}{*}{14.71} &\multirow{4}{*}{77.02} &\multirow{4}{*}{68.51} &\multirow{4}{*}{0.58} &$\rm{Tr2-0112}$     &7.28   &81.72   &43.77   &---\\
   &&&& &$\rm{Tr2-0121}$ &67.44  &75.53   &68.89   &---\\
   &&&& &$\rm{Tr2-0211}$ &23.41  &80.01   &80.56   &---\\
   &&&& &$\rm{Tr2-0212}$ &1.87   &74.87   &0.13    &31.19\\
\midrule
\multirow{7}{*}{$M=3$} &\multirow{7}{*}{0.80} &\multirow{7}{*}{67.17} &\multirow{7}{*}{58.79} &\multirow{7}{*}{1.51} &$\rm{Tr3-0113}$     &0.61  &64.29   &---    &---\\
   &&&& &$\rm{Tr3-0131}$  &82.05 &66.61   &65.53  &---\\
   &&&& &$\rm{Tr3-0212}$  &8.76  &69.28   &47.71  &9.15\\
   &&&& &$\rm{Tr3-0213}$  &1.19  &76.92   &---    &3.85\\
   &&&& &$\rm{Tr3-0311}$  &0.65  &12.50   &12.51  &---\\
   &&&& &$\rm{Tr3-0312}$  &6.53  &75.18   &11.68  &9.49\\
   &&&& &$\rm{Tr3-0313}$  &0.19  &80.00   &---    &20.00\\
\midrule
\multirow{4}{*}{$M=4$} &\multirow{4}{*}{5.69} &\multirow{4}{*}{87.03} &\multirow{4}{*}{15.80} &\multirow{4}{*}{2.34} &$\rm{Tr4-0122}$    &4.84  &71.81   &34.01  &41.06\\
   &&&& &$\rm{Tr4-0212}$ &1.52  &76.37   &54.43  &23.21\\
   &&&& &$\rm{Tr4-0221}$ &17.54 &68.67   &75.98  &---\\
   &&&& &$\rm{Tr4-0222}$ &74.75 &94.11   &---    &---\\
\midrule
$M=5$ &\sout{0.02} &--- &--- &--- &--- &--- &--- &--- &---\\
\midrule
\multirow{5}{*}{$M=6$} &\multirow{5}{*}{1.31} &\multirow{5}{*}{78.41} &\multirow{5}{*}{6.86} &\multirow{5}{*}{---} &$\rm{Tr6-0222}$     &20.01  &84.35   &---    &---\\
   &&&& &$\rm{Tr6-0223}$  &7.10   &91.22   &---    &---\\
   &&&& &$\rm{Tr6-0231}$  &9.90   &54.89   &69.29  &---\\
   &&&& &$\rm{Tr6-0232}$  &31.64  &84.85   &---    &---\\
   &&&& &$\rm{Tr6-0322}$  &26.41  &86.23   &70.29  &---\\
\midrule
Total &98.79 &79.35 &11.54 &0.23 &\multicolumn{5}{c}{---}\\
\bottomrule
\end{tabular}
\end{table*}

\clearpage

\end{document}